\documentstyle[preprint,epsf,aps]{revtex}
\tightenlines
\begin{document}
%\preprint{DFNAE-IF/UERJ-99/04}
\title{Pair of Heavy-Exotic-Quarks at LHC}
\author{ J.\ E.\ Cieza Montalvo $^*$} 
\author{P. P. de Queiroz Filho $^\dagger$}
\address{Instituto de F\'{\i}sica, Universidade do Estado do Rio de Janeiro\\ 
CEP $20559-900$ Rio de Janeiro, Brazil}
\maketitle
                  
\begin{abstract} 

We study the production and signatures of heavy exotic quarks pairs 
at LHC in the framework of the vector singlet model (VSM), vector doublet model (VDM) and fermion-mirror-fermion (FMF) model.  The  pair production cross sections for the electroweak and strong sector are computed.   

%We also exhibit some kinematical distributions.

%PACS number: 12.60.-i,13.85.Rm,14.80.-j

%
\vskip 2.5cm
%{\em Submitted to Phys.\ Rev.\ D}
%\end{center}
\end{abstract}

%\newpage

\section{Introduction}

One of the major problems  to be studied in particle physics concerns the spectrum of elementary fermions. Many models consider the possible existence of new generations of fermions, such  as composite  models \cite{af,bu1}, grand unified theories \cite{la}, technicolor models \cite{di}, superstring-inspired models \cite{e6}, mirror fermions  \cite{maa}, etc., which  predict the existence of new particles
with masses in turn of the scale of $1$ TeV. In this work, we will study the production mechanism for exotic-quarks  at LHC. It is assumed here that the mixing between ordinary and exotic leptons are of the same flavour.

The three models that we consider here include new fermionic  degrees of freedom, which introduce naturally a number of unknown 
 mixing angles and fermionic masses \cite{tom}. These 
models are: the vector singlet model (VSM) \cite{gon}, which consists 
in the inclusion of new left- and right-handed fermions in 
singlets, the vector doublet model (VDM) \cite{riz}, that can arise at low  energies in the ${\bf 27}$ representation of $E_{6}$ theories  
and the fermion-mirror-fermion (FMF) model
\cite{maa}, where the new particles are introduced to restore the
right-left symmetry.

Exotic fermions mixed with the standard fermions interact through the
standard weak vector bosons $W^{+}, W^{-}$ and $Z^{0}$, according to the Lagrangians \cite{sim1}

\begin{equation}
{\cal L}_{\rm NC} = \frac{g}{4 cos\theta_{W}} \left[{ \bar{F}_{i}}
\gamma^{\mu} (g_{V}^{ij} - g_{A}^{ij} \gamma^{5}) F_{j} +
{\bar{F}_{i}} \gamma^{\mu} (g_{V}^{ij} - g_{A}^{ij} \gamma^{5}) f_{j} \right]
Z_{\mu} 
\label{lag1}
\end{equation}
   
and 

\begin{equation}
{\cal L}_{\rm CC} = \frac{g}{2 \sqrt{2}} { \bar{Q_{i}} \gamma^{\mu} 
(C_{V}^{ij} - C_{A}^{ij} \gamma^{5}) Q_{j} W_{\mu} } ,
\label{lag2}
\end{equation}
where $F$ and $f$ are the exotic and standard fermions, $Q$ are the exotic quarks, $g_{V}^{ij}$ and $g_{A}^{ij}$ are the corresponding neutral
vector-axial coupling constants, and $C_{V}^{ij}$ and $C_{A}^{ij}$ are 
the charged vector-axial coupling constants, which can be obtained from Eq. (2) and are given in Table I, for each of the three models that we study here.

We consider here that all mixing angles have the value $\theta_{i}
= 0.1$, although phenomenological analysis \cite{tom1} give an upper
bound of $sin^{2} \theta_{i} \leq 0.03$. This means that the value of
$\theta_{i}$ can be scaled up to $0.173$.

\hskip 0.5cm

TABLE I. Coupling constants for a charged heavy fermion 
interaction: for the vector singlet model (VSM), the vector doublet model (VDM) and the fermion-mirror-fermion (FMF) model : 

\vskip 1cm

\begin{tabular}{|c|c|c|c|}  \hline\hline
Cou. & VSM & VDM & FMF \\ \hline  $C_{V}^{QQ^{'}}$ & $\sin \theta_{iL} \sin\theta_{jL}$ & $\cos (\theta_{iL} - \theta_{jL}) + 
\cos \theta_{iR} \cos \theta_{jR}$ & $\sin \theta_{iL} 
\sin \theta_{jL} + \cos \theta_{iR} \cos \theta_{jR}$ \\
\hline $C_{A}^{QQ^{'}}$ & $\sin \theta_{iL} \sin \theta_{jL}$
& $\cos (\theta_{iL} - \theta_{jL}) - \cos \theta_{iR} 
\cos \theta_{jR}$ & $\sin \theta_{iL} \sin \theta_{jL} - 
\cos \theta_{iR} \cos \theta_{jR}$ \\ \hline 
\end{tabular} 

\vskip 0.6cm

The other coupling constants for the neutral heavy fermion interaction are given in Ref. \cite{ciez1}.

%%%%%%%%%%%%%%%%%%%%%%%%%%%%%%%%

\section{Cross Section Production and Results}

The production mechanism can be studied through the analysis of the reaction $\bar{q} q (g g) \rightarrow \bar{Q} Q$, provided that there is enough available energy ($\sqrt{s} \geq 2M_{Q}$). Such process  takes  place through the exchange of a photon, a $Z^{0}$ and a $W$ in the $s$ channel, which concerns electroweak interactions,  and a  quark and a gluon in the $s$ and $t$ channel, corresponding to the strong  interaction.

For the $pp$ colliders, the contribution of the $Z$ and $W$ exchange, despite the high energy, is expected to be of the same order as the photon exchange, since the couplings of exotic quarks to $Z$ and $W$ have a weak interaction strength but the mixing angles are not so weak, while for the photon we have a electromagnetic strength only.

Using the interaction Lagrangians, Eqs. ($1$) and ($2$),  we first  evaluate the cross section for the electroweak case involving a  neutral  current and obtain:

\begin{eqnarray} 
\left (\frac{d \hat \sigma}{d\cos \theta} \right )_{\bar{Q} Q} =&&\frac{  \beta_{Q} \alpha^{2} \pi}{s}  \Biggl [\frac{c_{q}^{4}}{s^{2}}  \bigl [2 s M^{2}_{Q} + (M^{2}_{Q} - t)^{2} + (M^{2}_{Q} - u)^{2}   \bigr ]    \nonumber \\
&&+ \frac{1}{16 \sin^{4} \theta_{W} \cos^{4} \theta_{W}  (s -  M^{2}_{Z})^{2}} \bigl [2 s M^{2}_{Q} ( {g_{V}^{QQ}}^{2} - {g_{A}^{QQ}}^{2}) ( {g_{V}^{q}}^{2} + {g_{A}^{q}}^{2}) +  \nonumber \\   
&&({g_{V}^{QQ}}^{2} + {g_{A}^{QQ}}^{2}) ({g_{V}^{q}}^{2} + {g_{A}^{q}}^{2})   ((M^{2}_{Q} - t)^{2} + (M^{2}_{Q} - u)^{2})  \nonumber  \\ 
&&+ 4 (g_{V}^{QQ} g_{A}^{QQ} g_{V}^{q} g_{A}^{q} ((M^{2}_{Q} - u)^{2} - (M^{2}_{Q} - t)^{2}) \bigr ]  \nonumber \\
&&+\frac{c_{q}^{2}}{2 \sin^{2} \theta_{W} \cos^{2} \theta_{W} s (s -  M^{2}_{Z})}  \bigl [2 s M^{2}_{Q} ( g_{V}^{q} g_{V}^{QQ} +  (g_{V}^{q} g_{V}^{QQ} -  \nonumber  \\
&&g_{A}^{q} g_{A}^{QQ}) (M^{2}_{Q} - t)^{2} + (g_{V}^{q} g_{V}^{QQ} + g_{A}^{q} g_{A}^{QQ}) (M^{2}_{Q} - u)^{2}  \bigr ]   \Biggr ]   \;,
\end{eqnarray}
where $\beta_{Q} = \sqrt{1- 4 M_{Q}^{2}/s}$ is the velocity of the 
exotic-quark in the c.m. of the process, $c_{q}$ is the charge of the quark, $Q$ is the exotic quark and $\bar{Q}$ the exotic antiquark,  $M_{Z}$ is the mass of the $Z$ boson, $\sqrt{s}$ is the center of mass  energy of the $q \bar{q}$ system, $t = M_{L}^{2} - \frac{s}{2} (1 - \beta \cos \theta)$ and {} $u = M_{L}^{2} - \frac{s}{2} (1 + \beta \cos \theta)$,  where $\theta$ is the angle between the exotic quark and the incident quark, in the c.m. frame and the couplings $g_{V}^{QQ}$ and  $g_{A}^{QQ}$ are given in \cite{ciez1}.

For the charged current we obtain:

\begin{eqnarray} 
\left (\frac{d \hat \sigma}{d\cos \theta} \right )_{\bar{Q} Q}^{W}=&&\frac{ \beta_{Q} \alpha^{2} \pi}{32 \sin^{4} \theta_{W} s (s -  M^{2}_{W})^{2}} \bigl [2 s M^{2}_{Q} ({C_{V}^{QQ^{'}}}^{2} - {C_{A}^{QQ^{'}}}^{2}) +  \nonumber \\ &&({C_{V}^{QQ^{'}}}^{2} - {C_{A}^{QQ^{'}}}^{2}) (M^{2}_{Q} - t)^{2} + ({C_{V}^{QQ^{'}}}^{2} + {C_{A}^{QQ^{'}}}^{2}) (M^{2}_{Q} - u)^{2}  \bigr ]    \;,
\end{eqnarray}
where $M_{W}$ is the mass of the W boson  and $C_{V}^{QQ}$ and $C_{A}^{QQ}$ are given in Table I.

The total cross section for the process $pp \rightarrow qq \rightarrow 
\bar{Q} Q$ is related to the subprocess $qq \rightarrow  \bar{Q} Q$  total cross section $\hat{\sigma}$, by

\begin{equation}  
\sigma = \int_{\tau_{min}}^{1}
\int_{\ln \sqrt{\tau_{min}}}^{-\ln \sqrt{\tau_{min}}} d\tau \ dy \  
q(\sqrt{\tau}e^y, Q^2) q(\sqrt{\tau}e^{-y}, Q^2)  \hat{\sigma}(\tau, s)  \; , 
\end{equation}
where $\tau = \frac{\hat{s}}{s} (\tau_{min} = \frac{4 M_Q^2}{s})$, 
with $s$ being the center-of mass  energy  of the $pp$ system, and $q(x,Q^2)$ is the quark structure function.

The strong production of heavy standard quarks is very well  studied and those results can be used to the exotic quarks. With respect to the gluon fusion there is an additional enhanced process to be considered. This process is regulated  by  the axial  part of the Z boson and a Higgs. The exchange of a photon is not allowed by C conservation (Furry's theorem). The contribution of the Z boson is given by

\[
\left (\frac{d \hat{\sigma}}{d\cos \theta} \right )_{\bar{Q} Q}^{Z} =  \frac{{g_{A}^{Q}}^{2} N_{c} \alpha^2 \alpha_s^2}{256 \pi
\sin^4\theta_W} \frac{M_Q^2}{M_W^4} \beta_Q \left|
\sum_{q=u,d} T_3^q \left( 1 + 2 \lambda_q I_q  \right) \right|^2 
\]

and of a Higgs by 

\[
\left (\frac{d \hat{\sigma}}{d\cos \theta} \right )_{\bar{Q} Q}^{H} =  \frac{N_{c} \alpha^2 \alpha_s^2}{1024 \pi \sin^4\theta_W}  \frac{M_Q^2}{M_W^4}  \hat{s}^{2} \beta_Q^{3} \left| \chi (\hat{s})  \sum_{q=u,d}  \left(2 \lambda_q + \lambda_q (4 \lambda_q -1) I_q  \right) \right|^2  , 
\]
where the summations run over all generations. $T_{3}^{q}$ is the quark weak isospin [$T_{3}^{u(d)} = +(-)1/2$]. The loop function $I_{i} \equiv I(\lambda_{i} = m_{i}^{2} /\hat{s})$, is defined by 

\begin{eqnarray*}
I_i \equiv I_i (\lambda_i) = \int_0^1 \frac{dx}{x} 
\ln \left[1 - \frac{(1-x)x}{\lambda_i} \right] =   
\left \{ \begin{array}{l}
- 2 \left[ \sin^{-1}\left( \frac{1}{2 \sqrt{\lambda_{i}}} \right)\right]^2
\; , \;\;\; \lambda_i > \frac{1}{4} \\ \nonumber
\frac{1}{2} \ln^2 \left(\frac{r_+}{r_-}\right) - \frac{\pi^2}{2}  + i\pi 
\ln\left(\frac{r_+}{r_-}  \right)  \; , \;\;\; \lambda_i < \frac{1}{4} ,
\end{array}
\right.
\label{ii}
\end{eqnarray*}
with, $r_\pm = 1 \pm (1 - 4 \lambda_i)^{1/2}$ and $\lambda_i = 
m_i^2/\hat{s}$. Here, $i = q$ stands for the particle (quark ) running in the loop.

We have also defined 

\[
\chi (\hat{s}) = \frac{1}{\hat{s} - m_{H}^{2} + i M_{H} \Gamma_{H}} 
\]
with $\Gamma_{H}$ being the width of the Higgs boson.

The total cross section for the process $pp \rightarrow gg \rightarrow 
\bar{Q} Q$ is related to the subprocess $gg \rightarrow \bar{Q} Q$ total cross section $\hat{\sigma}$ by

\begin{equation}  
\sigma = \int_{\tau_{min}}^{1} 
\int_{\ln \sqrt{\tau_{min}}}^{-\ln \sqrt{\tau_{min}}} d\tau dy 
G(\sqrt{\tau}e^y, Q^2) G(\sqrt{\tau}e^{-y}, Q^2)  \hat{\sigma}(\tau, s)
\end{equation}
where $\tau = \frac{\hat{s}}{s} (\tau_{min} = \frac{4M_{\tilde{Q}}^2}{s})$, 
with $s$ being the center of mass  energy  of the $pp$ system, and $G(x,Q^2)$ is the gluon structure function.

In Fig. $1$ we show the cross sections for the production of 
exotic U-quarks, (for the D-quarks, the results for the cross section are   similar), $\bar{p} p \rightarrow \gamma, Z \rightarrow \bar{U}  U$,  we see  that for the three models that we are considering the cross  section are  similar one to another. In all numerical calculations we take $\sin^{2}  \theta_{W} = 0.2315$, $M_{Z} = 91.118$ GeV, $M_{W} = 80.33$  GeV and used the distribution functions given by M. Gl\"uck, E. Reya and A. Vogt  \cite{gluck}. Considering that the expected integrated luminosity for the LHC will  be of the order of $\propto 10^{5} pb^{-1}/yr$, and taking the mass of the quark-U equal to $500$ GeV, we expect to have a total of $\simeq  10^{3}$ exotic quark  pairs produced per year.

Fig. $2$ shows  the cross section for the production of exotic-U  quarks, $\bar{p} p \rightarrow  W  \rightarrow \bar{U}  U$, for the VDM and for the FMF model. For the quark mass of 500 GeV we expect a total of $\propto 10^{4}$ exotic quarks pairs produced per year for the VDM, while for the FMF we  estimate a total of $\propto 3.10^{3}$ events per year.

In Fig. $3$ we present the cross sections for the gluon fusion mechanism $\bar{p} p \rightarrow g g  \rightarrow \bar{U}  U$. For the same mass of 500 GeV we obtain, for the VDM model $\propto 10^{4}$ quark pairs produced per year and $\propto 3.10^{4}$ for the FMF model. In this case we take the mass of the Higgs equal to  $100$ GeV.

In Fig. $4$ we compare the cross sections as for the electroweak  case and as for the  strong case. We  observe that for large masses of exotic quarks, (more than $500$ GeV), the cross sections for the electroweak case begins to be competitive with the strong case, that is, for every $100$ events produced by the strong case we have $15$ to $20$ events produced by the electroweak case.

We make use of $p_{T}$ distributions to try to distinguish the signatures of the strong and electroweak cases. Fig. $5$  shows the  distribution of $p_{T}$ for the exotic quarks. We see that the exotic  quarks signal for the electroweak case is below for the QCD-case. We take in this case for the mass of the exotic quarks equal to $600$ GeV.

Fig. $6$ shows the distribution of transverse mass $M_{T} (qe, \nu)$. We see from the figure a Jacobian peak near the $M_{T} = M_{Q}$. The transverse mass is defined by \cite{barger}

\[
M_{T}^{2} (qe, \nu) = \left [(p_{qeT}^{2} + M_{qe}^{2} )^{1/2} + p_{\nu T} \right ]^{2} - (\vec{p}_{qeT} + \vec{p}_{\nu T} )^{2}
\]
where from here we know some information about the recoil q-jet. The $Q \bar{Q}$ events with real W decays are distinguished from the electroweak single and pair production of W's by the form of their decays, while for the heavy quark we have an odd number of quarks for the W we have an even number respectively.

It is necessary to point out that the process involving a boson Z to  gluon-gluon fusion for VSM and VDM-quarks gives a negligible contribution compared to the FMF-quarks, due to the small mixing angles.

%%%%%%%%%%%%%%%%%%%%%%%%%%%

\acknowledgments

I would like to thank to Prof. I. P. Calvalcante for a careful reading of the  manuscript and Prof. J. O. P. \'Eboli for some useful suggestions. One of the authors (P.P.Q.F.) would like to thank the Funda\c c\~ao de Amparo \`a Pesquisa do Estado do Rio de Janeiro for full financial support (contract No. E-26/150.639/99).

%%%%%%%%%%%%%%%%%%%%%%%%%%%

\newpage

\begin{center}
FIGURE CAPTIONS
\end{center}

\vspace{0.5cm}

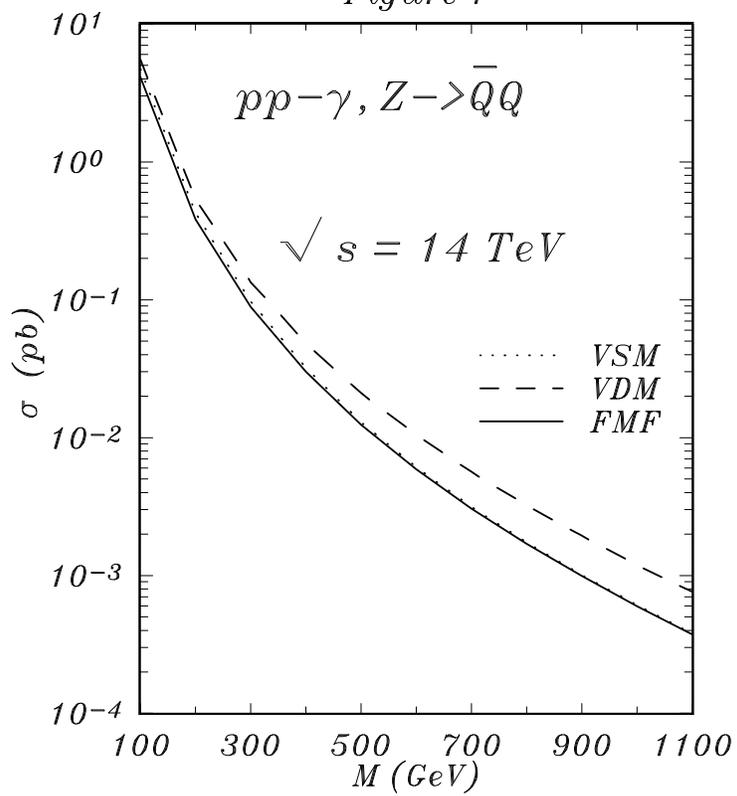
{\bf Figure 1}: Total cross section for the process $\bar{p} p \rightarrow \gamma, Z \rightarrow \bar{U}  U$, as a function of $M_{U}$ at $\sqrt{s} = 14$ TeV:  (a) vector singlet model (dotted line); (b) vector doublet model (dashed line); (c) fermion mirror fermion (solid line).

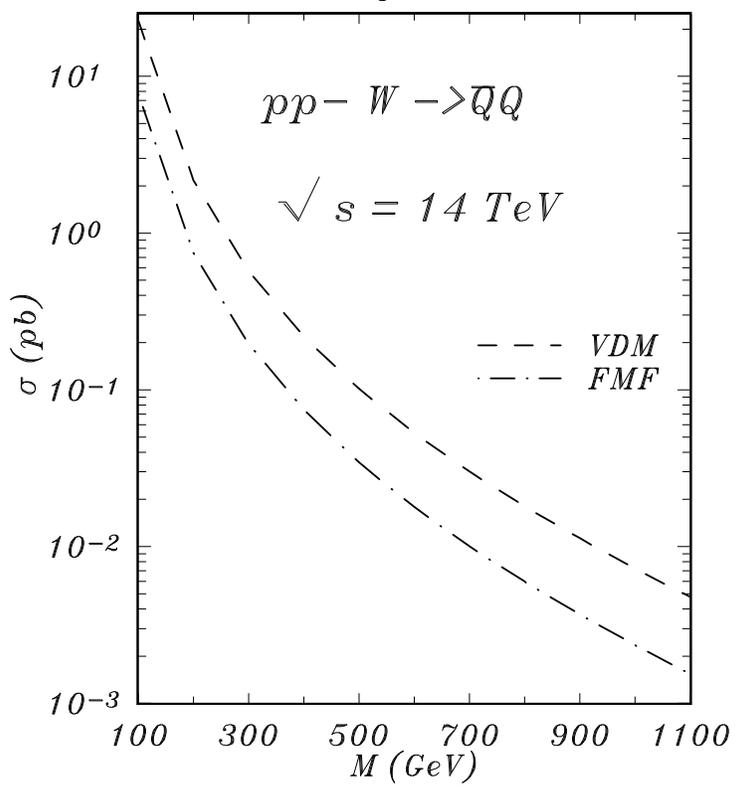
{\bf Figure 2}: Total cross section for the process $\bar{p} p \rightarrow  W  \rightarrow \bar{U} U$ as a function of $M_{U}$ at $\sqrt{s} = 14$ TeV:  (a) vector doublet model (dashed line); (b) fermion mirror fermion (dot dashed line).

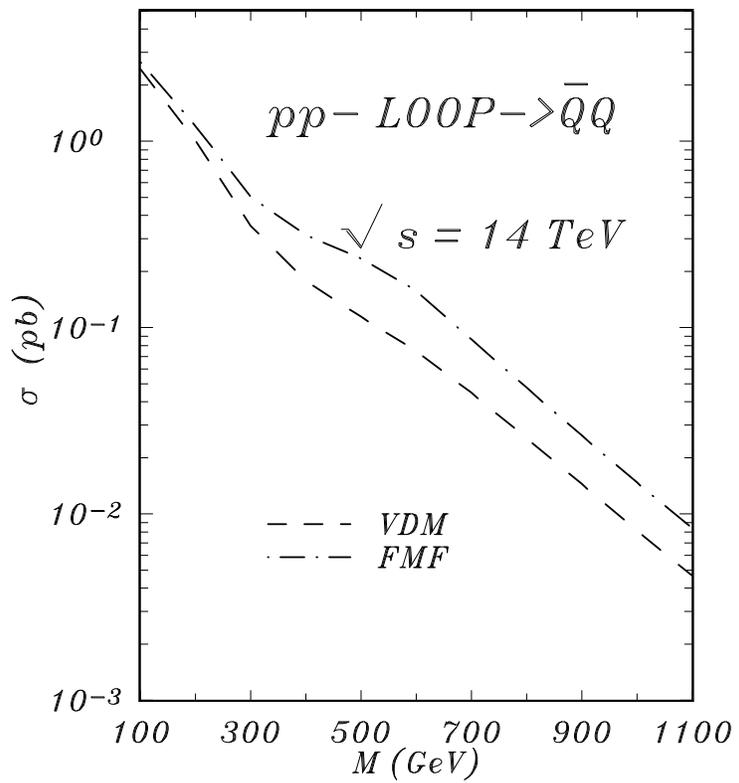
{\bf Figure 3}: Total cross section for the gluon-gluon fusion mechanism, $\bar{p} p \rightarrow g g  \rightarrow \bar{U} U$, as a function of $M_{U}$ at $\sqrt{s} = 14$ TeV:  (a) vector doublet model (dashed line); (b) fermion mirror fermion (dot dashed line).

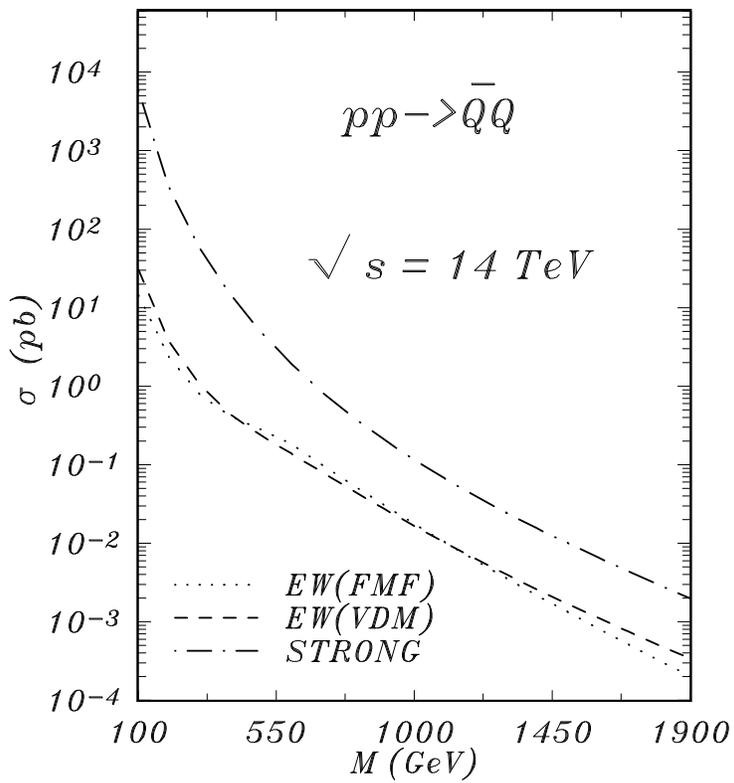
{\bf Figure 4}: Total cross section for the process 
$\bar{p} p \rightarrow \bar{U} U$ as a function of $M_{U}$ at $\sqrt{s} =
14$ TeV: (a) for the strong case (dot dashed line); for the electroweak  case: (b) FMF (doted line) and (c) VDM (dashed line).

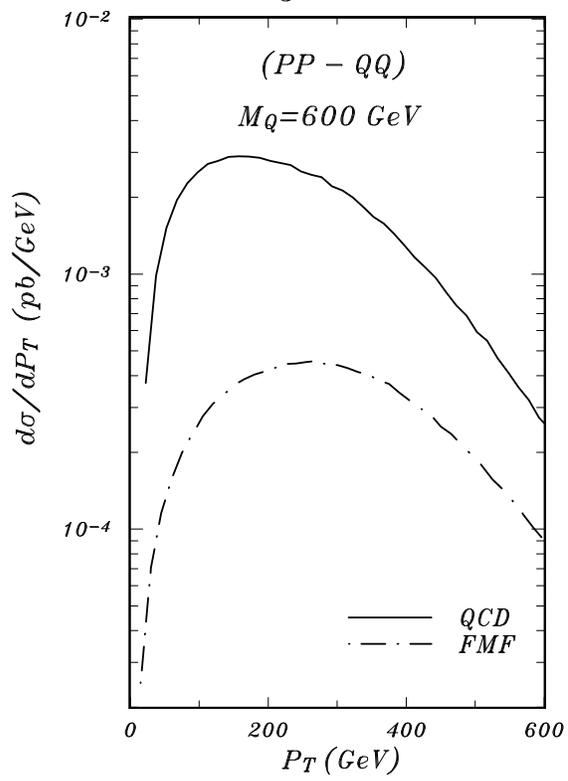
{\bf Figure 5}: $p_{T}$ distributions for $\bar{Q} Q$ production:  for exotic quarks mass equal to $600$ GeV.

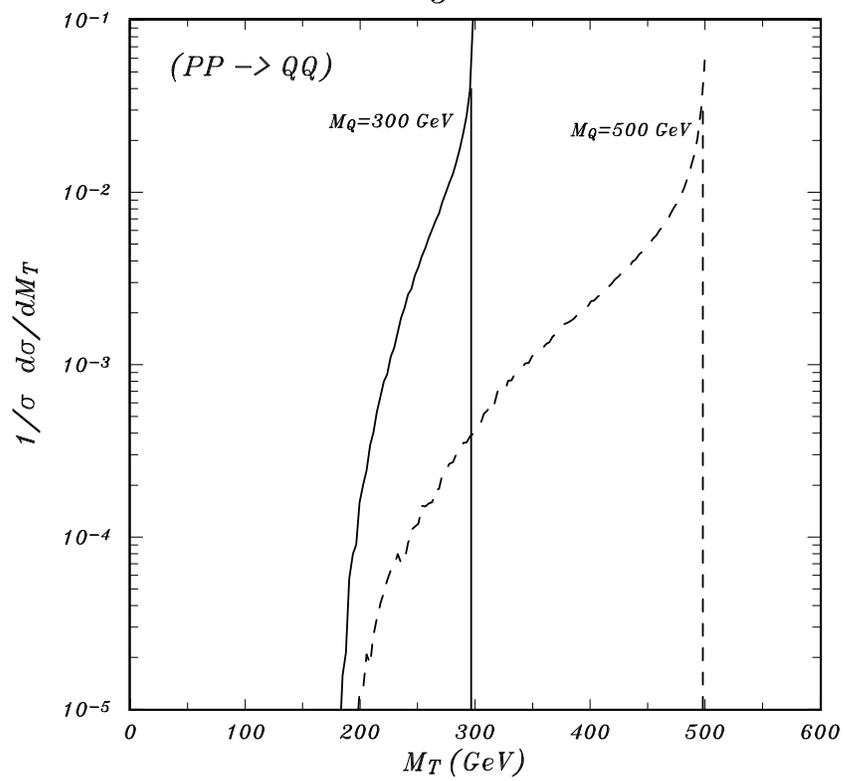
{\bf Figure 6}: Normalized distribution of transverse mass $M_{T} (qe, \nu)$ for  $\bar{Q} Q$ production: (a) for exotic quark mass equal to $300$ GeV (solid line); (b) for exotic quark mass equal to $500$ GeV (dashed line).

\end{document}